\documentclass[12pt]{article}
\usepackage{epsfig}

\newcommand\pubnumber{ZU-01/01}
\newcommand\pubdate{\today}

\newcommand\hepnumber{hep-ph/0101259}

\def\csumb{Institut f\"ur Theoretische Physik\\
Universit\"at Z\"urich\\
Winterthurerstrasse 190\\
8057 Z\"urich, Switzerland}

\def\Title#1{\begin{center} {\Large\bf #1 } \end{center}}
\def\Author#1{\begin{center}{ \sc #1} \end{center}}
\def\Address#1{\begin{center}{ \it #1} \end{center}}

\newcommand\pubblock{\rightline{\begin{tabular}{l} \pubnumber\\
         \pubdate\\ \hepnumber \end{tabular}}}
\newenvironment{Abstract}{\begin{quotation}  }{\end{quotation}}
\newenvironment{Presented}{\begin{quotation} \begin{center}
             Presented at the\end{center}
      \begin{center}\begin{large}}{\end{large}\end{center} \end{quotation}}
\def\Acknowledgments{\bigskip  \bigskip \begin{center}
          \large\bf Acknowledgments\end{center}}

\makeatletter
\def\section{\@startsection{section}{0}{\z@}{5.5ex plus .5ex minus
 1.5ex}{2.3ex plus .2ex}{\large\bf}}
\def\subsection{\@startsection{subsection}{1}{\z@}{3.5ex plus .5ex minus
 1.5ex}{1.3ex plus .2ex}{\normalsize\bf}}
\def\subsubsection{\@startsection{subsubsection}{2}{\z@}{-3.5ex plus
-1ex minus  -.2ex}{2.3ex plus .2ex}{\normalsize\sl}}

\renewcommand{\@makecaption}[2]{%
   \vskip 10pt
   \setbox\@tempboxa\hbox{\small #1: #2}
   \ifdim \wd\@tempboxa >\hsize     
       \small #1: #2\par          
     \else                        
       \hbox to\hsize{\hfil\box\@tempboxa\hfil}
   \fi}

 \def\citenum#1{{\def\@cite##1##2{##1}\cite{#1}}}

\newcount\@tempcntc
\def\@citex[#1]#2{\if@filesw\immediate\write\@auxout{\string\citation{#2}}\fi
  \@tempcnta\z@\@tempcntb\m@ne\def\@citea{}\@cite{\@for\@citeb:=#2\do
    {\@ifundefined
       {b@\@citeb}{\@citeo\@tempcntb\m@ne\@citea\def\@citea{,}{\bf ?}\@warning
       {Citation `\@citeb' on page \thepage \space undefined}}%
    {\setbox\z@\hbox{\global\@tempcntc0\csname b@\@citeb\endcsname\relax}%
     \ifnum\@tempcntc=\z@ \@citeo\@tempcntb\m@ne
       \@citea\def\@citea{,}\hbox{\csname b@\@citeb\endcsname}%
     \else
      \advance\@tempcntb\@ne
      \ifnum\@tempcntb=\@tempcntc
      \else\advance\@tempcntb\m@ne\@citeo
      \@tempcnta\@tempcntc\@tempcntb\@tempcntc\fi\fi}}\@citeo}{#1}}
\def\@citeo{\ifnum\@tempcnta>\@tempcntb\else\@citea\def\@citea{,}%
  \ifnum\@tempcnta=\@tempcntb\the\@tempcnta\else
  {\advance\@tempcnta\@ne\ifnum\@tempcnta=\@tempcntb \else\def\@citea{--}\fi
    \advance\@tempcnta\m@ne\the\@tempcnta\@citea\the\@tempcntb}\fi\fi}
\makeatother

\def\beq{\begin{equation}}
\def\eeq#1{\label{#1}\end{equation}}
\def\eeqn{\end{equation}}

\newenvironment{Eqnarray}%
   {\arraycolsep 0.14em\begin{eqnarray}}{\end{eqnarray}}
\def\beqa{\begin{Eqnarray}}
\def\eeqa#1{\label{#1}\end{Eqnarray}}
\def\eeqan{\end{Eqnarray}}

\let\bar=\overbar

\def\Dslash{\not{\hbox{\kern-4pt $D$}}}
\def\dslash{\not{\hbox{\kern-2pt $\del$}}}

\def\msb{{\bar{\ssstyle M \kern -1pt S}}}

\def\lsim{\mathrel{\raise.3ex\hbox{$<$\kern-.75em\lower1ex\hbox{$\sim$}}}}
\def\gsim{\mathrel{\raise.3ex\hbox{$>$\kern-.75em\lower1ex\hbox{$\sim$}}}}

\begin{document}
\begin{titlepage}
\pubblock

\vfill
\def\thefootnote{\fnsymbol{footnote}}
\Title{$CP$-Violation, the CKM Matrix and New Physics}
\vfill
\Author{Daniel Wyler}
\Address{\csumb}
\vfill
\begin{Abstract}
We discuss the influence of new physics on $CP$-violating observables.
Assuming the standard model gives a correct description of tree level
processes, we show how a consistent procedure can determine the
parameters of the standard model and check its validity also in loop
induced processes. A method to include new physics in a systematic
way is sketched.
\end{Abstract}
\vfill
\begin{Presented}
5th International Symposium on Radiative Corrections \\
(RADCOR--2000) \\[4pt]
Carmel CA, USA, 11--15 September, 2000
\end{Presented}
\vfill
\end{titlepage}
\def\thefootnote{\arabic{footnote}}
\setcounter{footnote}{0}

\section{Introduction}

Observation of novel phenomena often paves the way to new physics. For
instance, $\beta$
decays, parity and flavor violation required
the existence of a new force, the weak interactions.
At present, it is often thought that $CP$-violation could signal new
physics beyond the standard model. Although the latter can indeed
account for the observed effects \footnote{a notable exception is the
baryon asymmetry in the universe} (even $\epsilon'/\epsilon$
may be described by the standard model)
its predictions are not well tested (compared to physics
at LEP) and therefore a comprehensive study of $CP$-violation experiments
is important. As sketched in figure 1, $CP$-violation manifests itself in
many areas; only a comparison between them
can determine the correct description.
In the standard model, all $CP$-violation resides in the CKM matrix
\footnote{I do not discuss the so-called $\theta$ term} which
describes the couplings of the W-bosons to the quarks of different
charges. Therefore all appreciable $CP$-violation occurs
within flavor physics. Thus, one obvious strategy to search for new
forces and particles would be to look for non-zero $CP$-violating
effects where no flavour changes are involved, such as in electric dipole
moments or asymmetries in nuclear reactions. Unfortunately, the effects of
new physics are judged to be quite small (apart from the dipole
moments). Therefore more chance is given to the flavor sector instead,
that is the physics of
Kaons and mostly B-mesons. For a recent extensive review of $CP$-violation,
see ref. (\cite{nir}).

\begin{figure}[thb]
\begin{center}
\epsfig{file=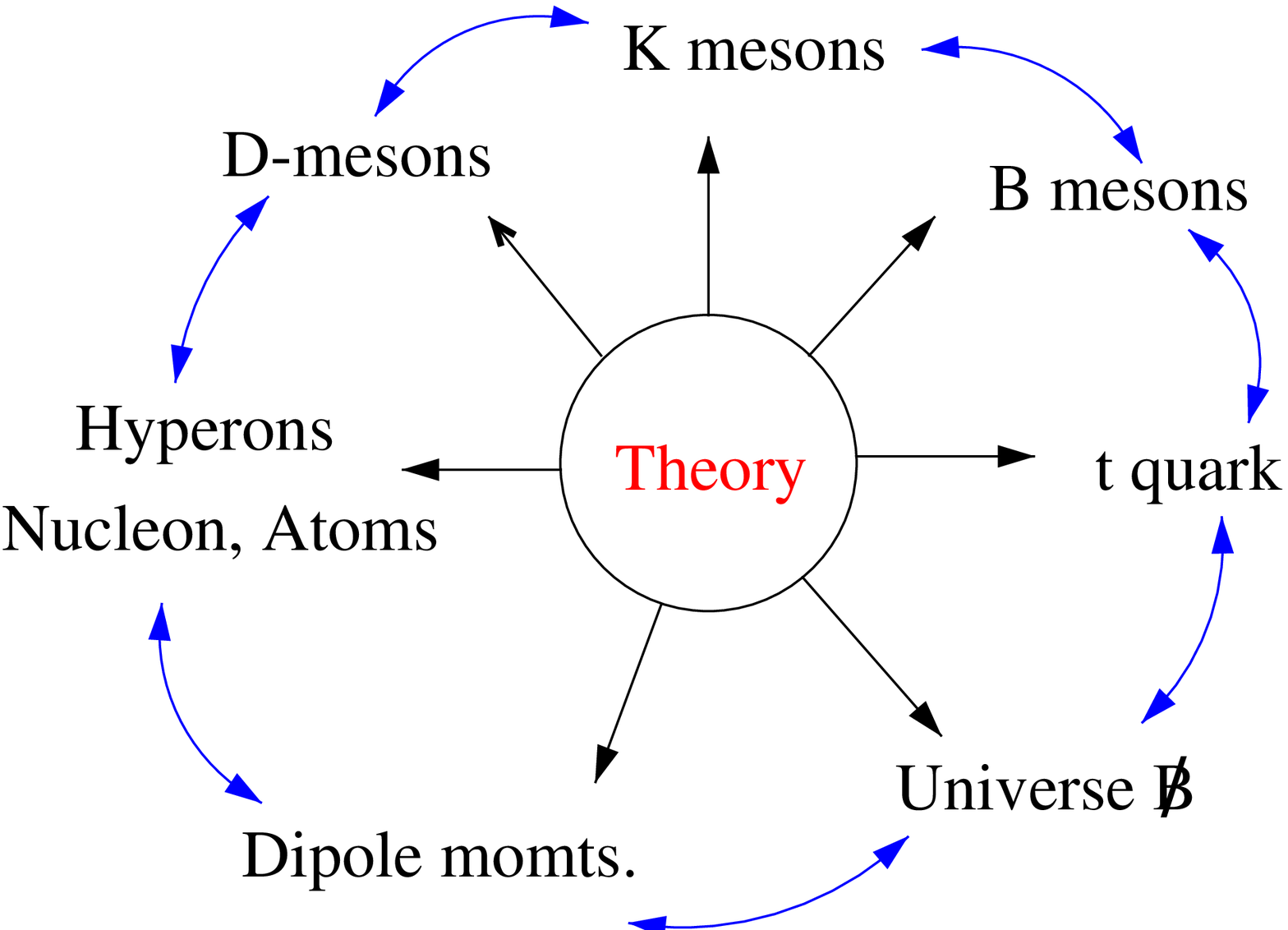, height=3in}
\caption[0]{\label{f1}CP-Violation}
\label{fig:general}
\end{center}
\end{figure}
The unitarity of the CKM matrix
\begin{equation}\label{2.67}
V =
\left(\begin{array}{ccc}
V_{ud}&V_{us}&V_{ub}\\
V_{cd}&V_{cs}&V_{cb}\\
V_{td}&V_{ts}&V_{tb}  
\end{array}\right)
\end{equation}
implies among others the triangle relation
\begin{equation}\label{2.87h}
V_{ud}^{}V_{ub}^* + V_{cd}^{}V_{cb}^* + V_{td}^{}V_{tb}^* =0
\end{equation}
which relates observable products of matrix elements and gives stringent
tests of the validity of the standard model. Using the Wolfenstein
parametrization and scaling as usual the bottom side to one,
we can write for the other sides of the scaled triangle
\begin{equation}\label{2.88b}
R_b = \frac{1}{A\lambda^3}V_{ud}^{}V_{ub}^*
=\bar\varrho+i\bar\eta
\qquad,
\qquad
R_t = \frac{1}{A\lambda^3}V_{td}^{}V_{tb}^*
=1-(\bar\varrho+i\bar\eta).
\end{equation}
Here, following ref. \cite{BLO}, the quantities
\begin{equation}
\bar\varrho = \varrho(1 - \frac{\lambda^2}{2})\,\,\,
\bar\eta = \eta(1 - \frac{\lambda^2}{2})
\end{equation}
are introduced to take into account even higher powers of $\lambda$.
\begin{figure}[b!]
\begin{center}
\epsfig{file=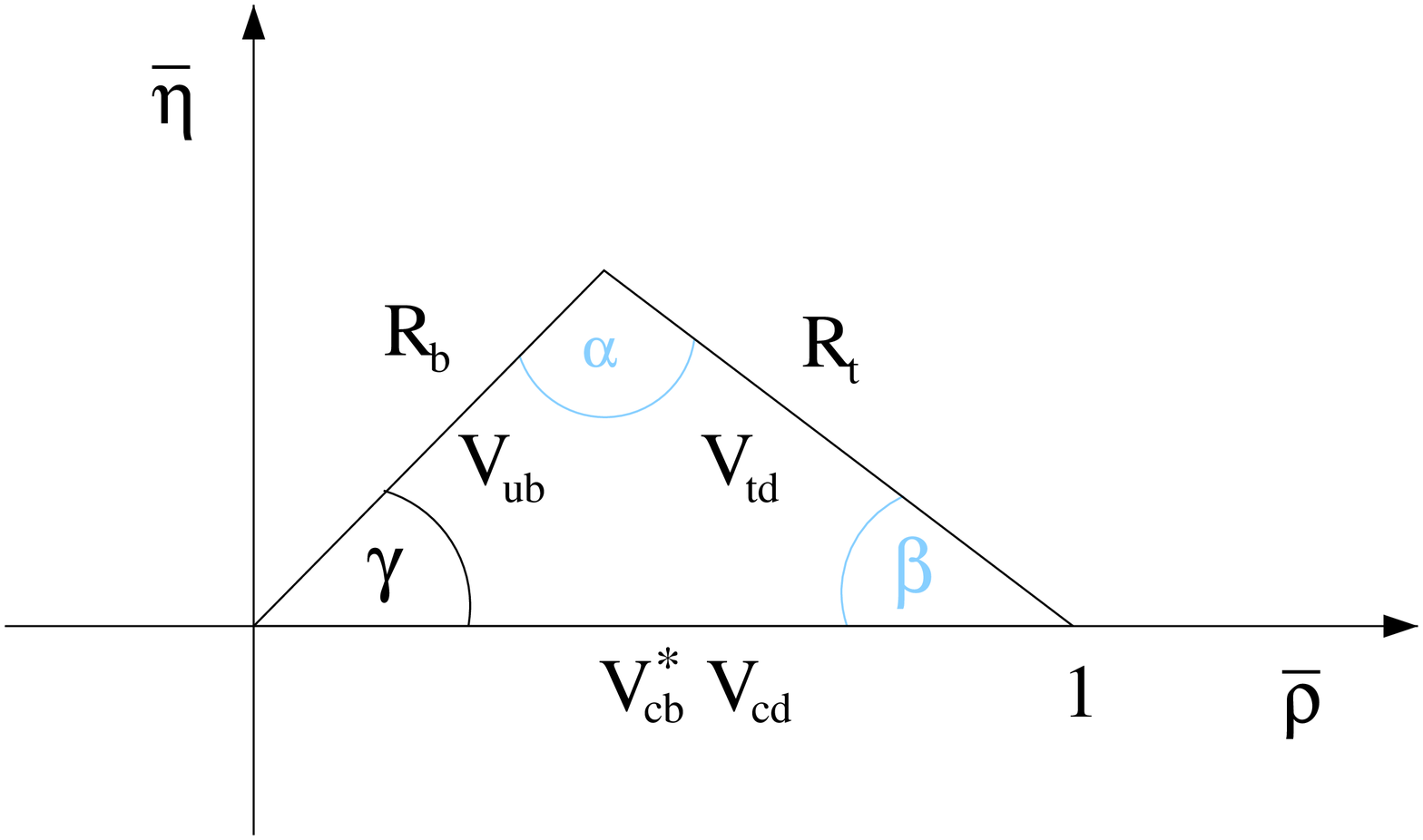,height=2.5in}
\caption[0]{\label{mhtanb}Unitarity triangle in the complex $(\bar\varrho,
\bar\eta)$ plane}
\label{fig:unitary}
\end{center}
\end{figure}

An elaborate analysis of superallowed $\beta$ decay, semileptonic
Kaon and $D$-meson decays and decays of $B$ mesons into charmed and charmless
final states yields \cite{pdg}
\begin{equation}
\begin{array}{cc}
V_{ud}= 0.9736 \pm 0.001&V_{cs}= 1.010 \pm 0.16\\
V_{us}= 0.2205 \pm 0.0018&V_{cd}= 0.224\pm 0.016\\
V_{ub}= 0.04  \pm 0.002&V_{cb}= 0.0036\pm 0.006\\
\end{array}
\end{equation}

These are (apart from corrections) all tree-level processes and therefore
thought to be governed by the standard model \footnote{of course, the small
$b \to u$ transition could be due to new physics}. They are however not
sufficient to check unitarity (unless very precise data from $t$ decays
would be available, or if the sum of the squares would be significantly
away from $1$).

Further input comes from loop-induced observables. They can be
calculated within perturbation theory and input from hadronic
physics. While the former are rather reliable and usually give
results accurate to 10 percent or so, the latter are generally
difficult to estimate. One usually considers the Kaon-mixing
quantity $\epsilon_K$, the mass difference of the $B$ and the
$\bar B$ mesons (and also of the $B_s$ and $\bar B_s$ mesons).
This analysis has resulted in the range of values for the three
angles $\alpha$, $\beta$ and $\gamma$ of the unitary
triangle and its sides.
The hadronic uncertainties are summarized in \cite{hadro} and are
reflected by
\begin{equation}
|R_b| = 0.39 \pm 0.07~~~~~~|R_t| = 0.98 +0.04 -0.22
\end{equation}
and by \cite{Stocchi,ALI00,SCHUNE}
\begin{equation}
\label{sinth}
(\sin 2\beta)_{\rm SM} =
0.75\pm0.20.
\end{equation}
The new results of last summer and of the beginning of this year
concern the angle $\beta$. It was found that
the coefficient $a$ of $sin(\Delta M_{B_d})$ in the asymmetry for $B \to
J/\Psi K_S$ is
\begin{eqnarray}
a = 0.79 \pm 0.4 (CDF)\cite{CDFB}\\
a = 0.58 \pm 0.35 (Belle)\cite{Belle}\\
a = 0.34 \pm 0.25 (BaBar)\cite{BaBar}
\end{eqnarray}

In the standard model, one has $a=sin(2\beta)$; comparing
eqs. (\ref{sinth}) and (10) we see a surprising inconsistency.
Of course, this is a preliminary
result, and may disappear as experiments collect more statistics.
However, it makes it mandatory to investigate $CP$-violation in
a (standard) model independent way. Unless $CP$-violation within the
standard model is grossly wrong, this program essentially amounts
to making many measurements and extracting discrepancies between
quantities thought to be the same in the standard model. Many
authors have discussed this situation; see e.g. \cite{NIR00,SW,NK00,XING,bubu}.

\section{A more general framework}

New physics may affect every process. Because the standard model
describes the most important weak decays, we will assume that it
accounts for semileptonic and tree-level quark
decays, at least to the required accuracy. This assumption can
be tested, by investigating the consistency of different semileptonic
decays, bounds from LEP etc. As an example consider the strengths of the
effective Hamiltonians
\begin{eqnarray}
\label{forf}
{\cal H}_{eff} =
G_F (\bar c_L \gamma_{\mu}b_L)(\bar s_L \gamma^{\mu}c_L)\\
{\cal H}_{eff} =
G_F (\bar u_L \gamma_{\mu}b_L)(\bar s_L \gamma^{\mu}u_L).
\end{eqnarray}
In the standard model, they are proportional to
$\lambda^2$ and $\lambda^4$, respectively. On the other hand, a new
neutral intermediate boson, say $Z'$, may exist, coupled to the currents
$(\bar s_L \gamma_{\mu}b_L)$ and $(\bar c_L \gamma_{\mu}c_L)$.
If it also couples to quark and lepton pairs, such as
$(\bar u_L \gamma_{\mu}u_L)$ and $(\bar c_L \gamma_{\mu}c_L)$, it would
contribute to the above Hamiltonians, to $B_s$ mixing,
to $B_s \to l^+ l^-$ etc. If the couplings are the same for all
these pairs, the effective strength would be the same for the two
terms in eqs. (\ref{forf}) and (12).
Therefore a new $Z'$-mediated interaction would induce a deviation
from the standard model result that the couplings of the two
interactions have a relative strength of $\lambda^2$.
Thus detailed studies could in principle also test the first assumption.
But of course, there are various experimental and theoretical
difficulties to overcome before one will obtain accurate enough results.
\begin{figure}[b!]
\begin{center}
\epsfig{file=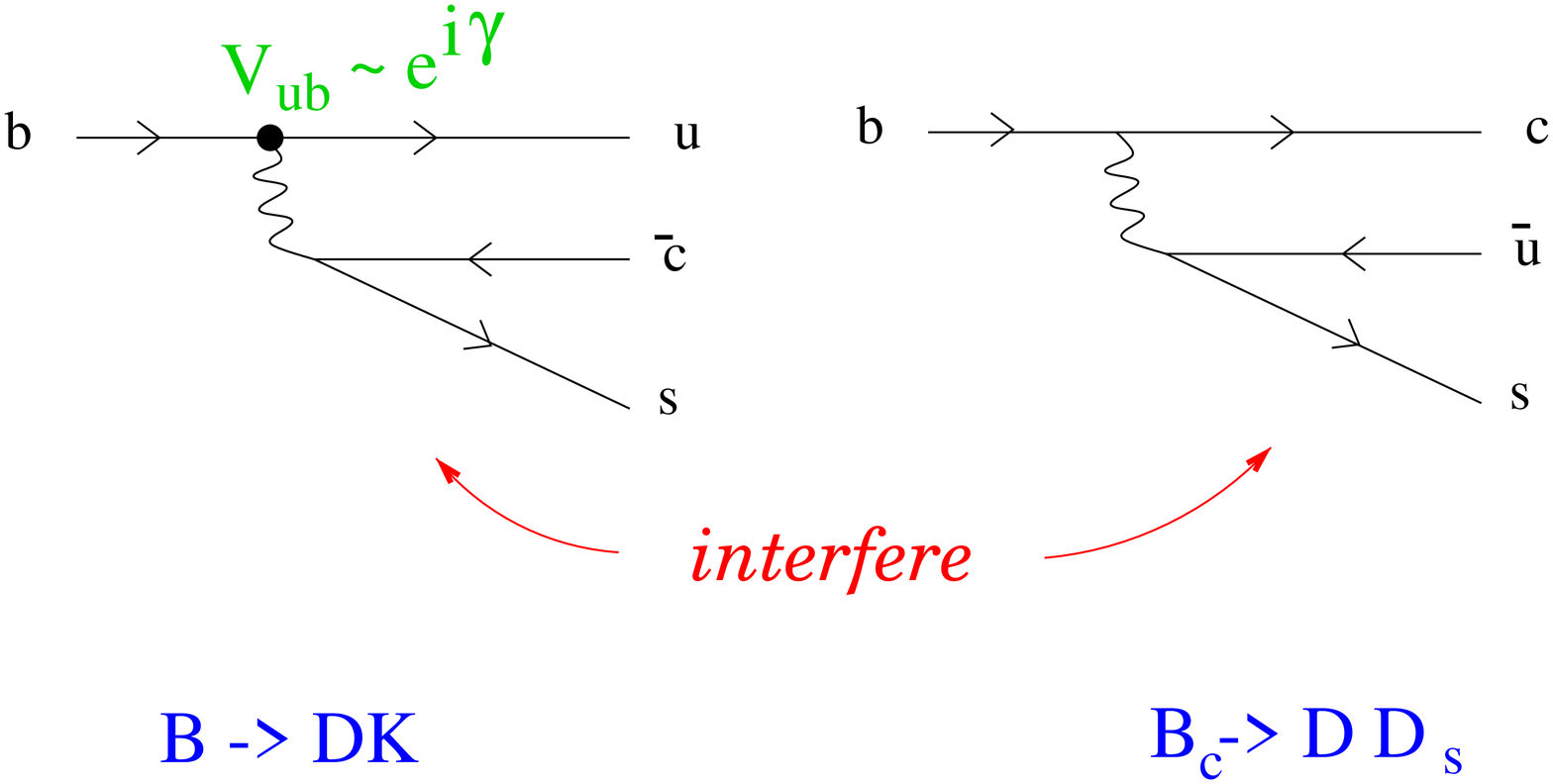, height=3in}
\caption[0]{\label{f2}two quark diagrams whose interference gives $\gamma$}
\label{fig:interference}
\end{center}
\end{figure}

From fig. 3 we see that the determination of the angle $\gamma$ from
tree level processes involves the interference of amplitudes proportional
to $V_{ub}$ and $V_{ub}$ respectively. This is achieved in processes
where the two diagrams of fig. 3 contribute. A well known example are
the decays $B \to D K$ \cite{gw,ads}; more recently the advantage
of $B_c \to D D_s$
was stressed \cite{flwy}. The idea is the same as in the previous
papers on $B \to D K$ : One needs to measure the six
amplitudes shown in Fig. 4. However due to the different $CKM$ elements,
the sides of the triangles in Fig. 3 are now 
of similar length and an extraction of
$\gamma$ seems possible with the $10^{10}$ or so $B_c$-mesons expected
at LHC. This method does not suffer from hadronic uncertainties.
\begin{figure}[b!]
\begin{center}
\epsfig{file=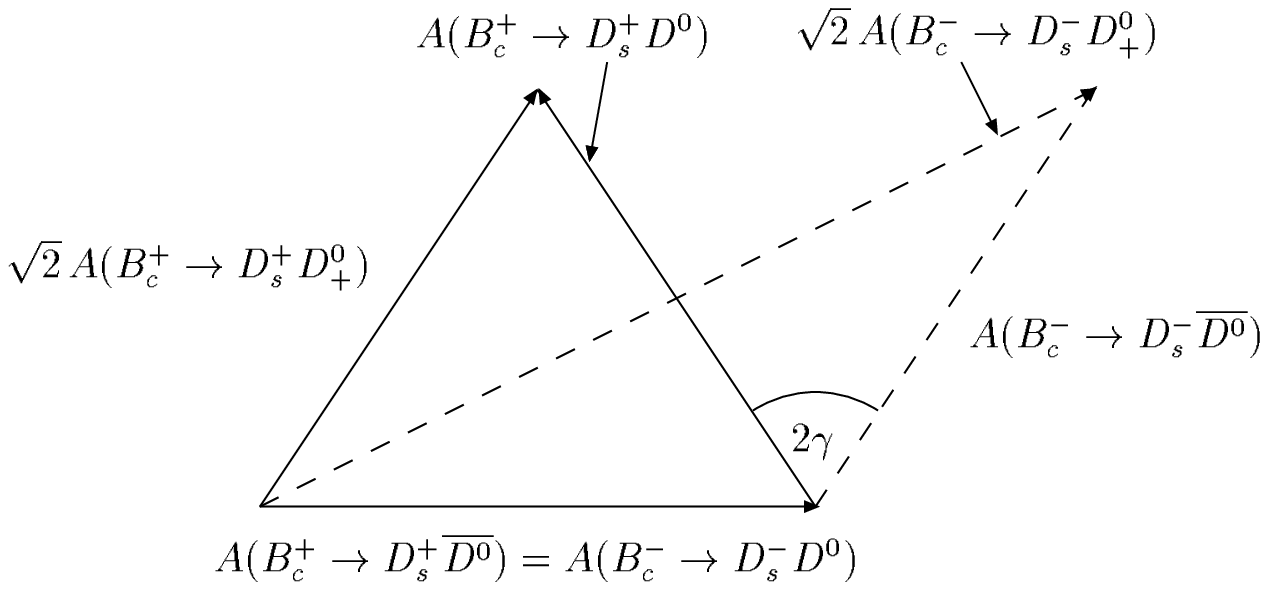,height=3in}
\caption[0]{\label{f3}The extraction of $\gamma$ from
$B_c^\pm\to D^\pm_s\{D^0,\overline{D^0},D^0_+\}$ decays}
\label{fig:triangles}
\end{center}
\end{figure}

The experimental difficulties associated with these decays
have lead to other possiblities. The decays $B \to K \pi$ are
sensitive to the interference
of the tree level diagram (with $V_{ub}$) and the penguin diagram. This
also yields
the angle $\gamma$ if the penguin graph has no extra phase.
This decay has been discussed by many people \cite{mf}.

A third possibility that was investigated are the decays $B^0 \to D^{\pm}
\pi^{\mp}$ \cite{fleisch}. The usual mixing-decay formalism yields
for the time dependent asymmetries the coefficients
\begin{eqnarray}
a \sim Im(e^{-i(2 \phi_{mix} +\gamma)})const\\
\bar a ~ \sim Im(e^{-i(2 \phi_{mix} +\gamma)})/const.
\end{eqnarray}
where const is an unknown hadronic number. It cancels in the product
which then yields the combination
\begin{equation}
2 \phi_{mix} +\gamma.
\end{equation}
The $B \bar B$ mixing angle $\phi_{mix}$ can be determined as usual from
the decay $B \to J/\Psi K_s$.

The other angles of the triangle cannot be determined independently
by a tree
level analysis. But we see, that the tree level analysis allows
to determine the unitary triangle of the standard model. It yields,
in principle, also the unknown
side $R_t$
and the angle $\beta$. Any further independent measurement of these
quantities checks the standard model with high accuracy, but it requires
loop effects.

\section{New Physics: Phenomenology}

Among the $CP$-violating observables, the mixing-decay asymmetry is
the cleanest theoretically \cite{bs}. It is therefore reasonable to start an
investigation of new physics with this quantity. Denoting the coefficient
of $sin(\Delta m t)$ by $a$, one has in general
\begin{equation}
a_{M \to F} = Im((\frac{p}{q})_M\frac{a}{\bar a}(\frac{p}{q})_F)
\end{equation}
where ($\frac{p}{q}$) are the mixing parameters and $a$, $\bar a$ the
amplitudes for $M \to F$ and $M \to \bar F$, respectively.

Setting for the $B$-meson mixing element $M_{12}$
\begin{equation}
M_{12} = r^2 e^{2 i \phi^{NP}}e^{2 i \beta}|M_{12}^{SM}|
\end{equation}
to account for a possible new phase and magnitude of the mixing,
the asymmetry coefficient is given in the table below:

\begin{center}
\begin{tabular}{|c|c|c|c|c|} \hline
$quarks$&$B_d$&$a$&$B_s$&$a$\\ \hline
$b \to c \bar c s$&$\Psi K_s$&$\beta + \phi^{NP}_d$&$D D_s$&$\phi^{NP}_s$\\
$b \to s \bar s s$&$\Phi K_s$&$\beta + \phi^{NP}_d + \phi^{A}$&$\Phi \Phi$&$\phi^{NP}_s + \phi^{A}$\\
$b \to u \bar u s$&$\pi \pi$&&&\\
$b \to c \bar c d$&$D^+ D^-$&&&\\
$b \to u \bar u s$&$\pi^0 K_s$&&&\\
$b\to s \bar s s$&$\Phi \pi$&&&\\ \hline
\end{tabular}
\end{center}
The phase $\phi^A$ takes into account a possible new phase in the decay.
The entries left out receive possibly sizeable contributions from
penguin diagrams and cannot be brought to the simple form. This result
tells us that comparing the different asymmetries, we can check the
consistency of the standard model and determine the phases of new physics.

New physics will also influence other $CP$-violating observables, such
as the  direct asymmetries of, say, charged B-meson decays. In cases
such as $B \to K \pi$, where the asymmetry is small in the standard model
new physics may give rise to sizeable asymmetries. Of course,
one needs to continue the experimental search for these, but because of the
difficulty of calculating direct asymmetries, only quantitative
statements are possible.

\section{New Physics: Analysis}

If new physics is associated with a scale $\Lambda$ much above
the weak scale ($\sim M_W$), the total Lagrangian density may be
written in the form \cite{bw}
\begin{equation}
{\cal L} = {\cal L}^{SM} + \sum {d_i {\cal O}_i^{NP}}
\end{equation}

where the $O_i$ are operators of dimension six induced by new physics
and their coefficients $d_i$ are of order($1/{\Lambda^2}$).
This 'effective'
Lagrangian is not renormalizable, and therefore one usually uses the
new operators only at tree level (see a discussion by). The $CP$-
violation induced by effective operators ${\cal O}_i^NP$ can in most
cases only be seen
when they are in loops, because the imaginary part (discontinuity)
of the corresponding Feynman graph is responsible for $CP$-asymmetry.
\footnote{an exception is the electric dipole moment} At low energies
we then have an effective Hamiltonian
\begin{equation}
{\cal H} =\sum {c_i{\cal O}_i^{SM}} + \sum {d_i {\cal O}_i^{NP}}.
\end{equation}

The amplitudes for a process $I \to F$ and the CP conjugated
one $\bar I \to \bar F$ then are
\begin{equation}
A(I \to F) =\sum {c_j (R_j + iI_j)^{SM}}  + \sum {d_j (R_j + iI_j)^{NP}}
\end{equation}
where $R$ and $I$ are the dispersive and absorptive parts of the
matrix elements. For the charge-conjugated process we have similarly
\begin{equation}
A(\bar I \to \bar F) =\sum {c_j^* (R_j + iI_j)^{SM}}  + \sum {d_j^* (R_j + iI_j)^{NP}}
\end{equation}
When we calculate the $CP$-violating asymmetry $\alpha \sim (|A(I \to F)|^2-
|A(\bar I \to \bar F)|^2)$, we obtain in leading order in QCD and in NP
\begin{equation}
\alpha \sim Im(cd^*)(R^{SM}I^{NP}-R^{NP}I^{SM}).
\end{equation}
$R^{NP}$ is a (finite) tree level amplitude, however also the loop
$ I^{NP}$ is finite.
Therefore the problems associated with a the non-rnormalizable theory
$\sum {d_i {\cal O}_i^{NP}}$ disappear and exact predictions are indeed
possible for the the $CP$-violating asymmetry. Therfore, an analysis
of the effects of new operators is possible also at for CP-violating
asymmetries, and not just at tree level!

\section{New Physics: Models}

Virtually any model beyond the standard one carries new sources for
flavour and $CP$-violations. It is therefore more economical to
look at them in increasing complexity.

The simplest one are the minimal flavour violating ones (MFV) where
all sources of flavour violation reside in the CKM matrix. This
results in many cases in a simple modification of the
coefficients in the usual loop expressions. However, there still
is a unitary triangle, but its sizes and angles may change. It was analyzed by
Ali and London \cite{ALI00}; recently Buras and Buras \cite{bubu} found a
clever lower bound on $sin(2\beta)$. The idea is simple. For both
$\epsilon$ and the $B$-meson mass difference, the standard model
contribution consists mostly of a $W-W-t-t$ box diagram; its value
might be denoted by $F_{tt}$. The $MFV$ modify this to
\begin{equation}
F_{tt}=S_0(m_t)~ (1+f)~.
\end{equation}
Then we can write for $\epsilon_K$
\begin{equation}\label{100}
\epsilon_K  \sim \bar\eta \left[(1-\bar\varrho) A^2 \eta_2 F_{tt}
+ P_c(\varepsilon) \right] A^2 \hat B_K
\end{equation}
while the $B$-meson mass difference yields the relation
\begin{equation}\label{RT}
R_t= 1.26~ \frac{ R_0}{A}\frac{1}{\sqrt{F_{tt}}}~,
\end{equation}
where
\begin{equation}\label{R0}
 R_0= \sqrt{\frac{(\Delta M)_d}{0.47/{\rm ps}}}
          \left[\frac{200~mev}{F_{B_d} \sqrt{\hat B_d}}\right]
          \sqrt{\frac{0.55}{\eta_B}}~.
\end{equation}

With
\begin{equation}
\label{ss}
\sin 2\beta=\frac{2\bar\eta(1-\bar\varrho)}{R^2_t}
\end{equation}
one gets \cite{BLO}
\begin{equation}
\label{main}
\sin 2\beta=\frac{1.26}{ R^2_0\eta_2}
\left[\frac{0.226}{A^2 \hat B_K}-\bar\eta P_c(\varepsilon)\right].
\end{equation}
Since unitarity implies $\bar\eta \le R_b~$, there exists a lower bound
on $\sin 2\beta$. A careful numerical analysis implies \cite{newbur}
\begin{equation}
\sin 2\beta \ge 0.42.
\end{equation}
The lower bound in fact corresponds to a $F_{tt}$ which is three times
larger than the standard model value.

Supersymmetry is a attractive candidate for new physics. In general, there
are many new $CP$-violating phases. Since they can directly affect
observables such a the electric dipole moment, it is natural
to take them to be small (approximate $CP$-violation, \cite{nir}). In
this situation, also $CP$-violating effects in the $B$-system are small.
This implies a small angle $\beta$.
This is in contrast to the standard model, where the flavour structure
suppresses CP-violation.
The problem with this scheme is that it is hard to get $\epsilon_K$ right
and that $\epsilon'/\epsilon$ tends to be to small.

Similarly, models with left-right symmetry tend to have small $CP$-violating
phases, thus the effects tend to be small also.

\section{$CP$-violation in $D$-mesons}

In the standard model, $CP$-violation is small in the $D$-System. This
is partly due to the rather large tree-level decay rates and small
coupling of the third generation. Therefore one would expect new
physics $CP$-violation mostly in the mixing (see \cite{nir} for a
more detailed discussion).
Recent studies of time-dependent decay rates of $D^0\rightarrow K^+\pi^-$
by the CLEO collaboration \cite{cleo} and measurements of the combination of
$D^0\rightarrow K^+K^-$ and $D^0\rightarrow K^-\pi^+$ rates by the FOCUS
collaboration \cite{focus} gave first information on the mixing.

As usual, one define the mixing quantities
\begin{equation}\label{DelMG}
x\equiv{m_2-m_1\over\Gamma},\ \ \
y\equiv{\Gamma_2-\Gamma_1\over2\Gamma}.
\end{equation}
CP-violation in the mixing is defined by the angle $\phi$. The experiments
find that the quantity $y\cos\phi$ is significantly larger than the
expectation in the standard model. The errors being large, this result
is not yet significant, but it shows the potential of $D$-meson physics.

\section{$K$-Physics}

Finally let me mention $K$-physics. Of course, efforts continue in
calculating $\epsilon'/\epsilon$ and to overcome the hadronic
difficulties, and there will be substantial progress.
However, the rare decays $K^+ \to \pi^+ \nu \bar \nu$ and
$K^0 \to \pi^0 \nu \bar \nu$ provide a theoretically
clean way to measure (in the standard model) $|V_{td}|$ and $Im V_{td}$
\cite{kburas}.
Clearly, this can be used as a test of the unitary triangle, however
the measurement of the neutral decays is not easy and probably many years
away.

\section{Conclusions}

The new results on $\sin 2\beta$ are surprising; they may indicate
a failure of the standard model. Several parameters have to be stretched
beyond their reasonable values to account for them. One can
modify the standard model to accommodate the small value of$\sin 2\beta$,
but it is not clear that these modifications are consistent.

Nevertheless, the result brings back the (old) view, that a (standard)
model independent and broad analysis of $CP$-violation
is required in order to fully understand this phenomenon and the
need for new interaction. this implies in particular measurements of
many decay channels.

I have sketched strategies to determine the source of $CP$-violation for
the case that the standard model accounts for tree level processes
and given a phenomenological framework to calculate the effects
of new operators. Needless to say that all of this will take many years
of hard work on both the experimental and the theoretical side and
that also less perfect measurements have to be pursued.

\Acknowledgments
I thank H. Haber and all speakers for an interesting and exciting
meeting. I am grateful to Georgia Hamel and Jacqueline Pizzuti at SCIPP and
Ruth McDunn and Terry Anderson at SLAC for their tireless efforts in
helping to make RADCOR-2000 a success. I thank my collaborators,
expecially G. Colangelo and R. Fleischer.

\end{document}